\documentclass[aps,prl,preprintnumbers,amsmath,amssymb,latexsym,nofootinbib,array,enumerate,letter,twocolumn,superscriptaddress]{revtex4-1}
\usepackage{cancel}
\usepackage{here}
\usepackage{comment,braket}
\usepackage{epsf}
\usepackage{amsmath}
\usepackage{graphics}
\usepackage{amsfonts}
\usepackage{amssymb}
\usepackage{latexsym}
\usepackage{color}
\usepackage{ulem}
\input{colordvi.tex}
\usepackage{feynmp}
\usepackage[pdftex]{graphicx}
\usepackage{bm}
\usepackage[pdftex]{hyperref}
\usepackage[svgnames]{xcolor}
\definecolor{phthaloblue}{rgb}{0.0, 0.06, 0.54}
\hypersetup{
    colorlinks=true,
    linkcolor=phthaloblue,
    citecolor=blue,
    filecolor=blue,
    urlcolor=phthaloblue,
    }

\newcommand{\beq}{\begin{eqnarray}} 
\newcommand{\eeq}{\end{eqnarray}}

\def\({\left(}
\def\){\right)}
\def\[{\left[}
\def\]{\right]}

\def\feff{f_{\rm eff}}
\def\veff{V_{\rm eff}}
\def\vpq{V_{\rm PQ}}

\newcommand{\bel}[1] {\begin{equation}\label{#1}}
\newcommand{\beal}[1] {\begin{eqnarray}\label{#1}}
\newcommand{\be}{\begin{equation}}
\newcommand{\ee}{\end{equation}}
\newcommand{\bea}{\begin{array}} 
\newcommand{\eea}{\end{array}}

\newcommand{\MeV}{\  {\rm MeV} }
\newcommand{\GeV}{\  {\rm GeV} }
\newcommand{\TeV}{\  {\rm TeV} }

\begin{document}
%\preprint{RESCEU-1/16}

\title{Axiogenesis}

\author{Raymond T. Co}
\affiliation{Leinweber Center for Theoretical Physics, University of Michigan, Ann Arbor, MI 48109, USA}
\author{Keisuke Harigaya}
\affiliation{School of Natural Sciences, Institute for Advanced Study, Princeton, NJ 08540, USA}

\date{\today}

\begin{abstract}
We propose a mechanism called axiogenesis where the cosmological excess of baryons over antibaryons is generated from the rotation of the QCD axion.
The Peccei-Quinn (PQ) symmetry may be explicitly broken in the early universe, inducing the rotation of a PQ charged scalar field. The rotation corresponds to the asymmetry of the PQ charge, which is converted into the baryon asymmetry via QCD and electroweak sphaleron transitions. In the concrete model we explore, interesting phenomenology arises due to the prediction of a small decay constant and the connections with new physics at the LHC and future colliders and with axion dark matter.
\end{abstract}

\preprint{LCTP-19-27}

\maketitle

{\bf Introduction.}---%
One of the goals of fundamental physics is to understand the origin of the Universe.
For this purpose, the Standard Model (SM) of particle physics needs an extension to explain the cosmological excess of matter over antimatter. Mechanisms to generate the baryon asymmetry have been intensively studied in the literature under the name of baryogenesis.
The proposed origins of the baryon asymmetry include explicit baryon or lepton number violation from a) the supersymmetric partners of baryons or leptons in the Affleck-Dine mechanism~\cite{Affleck:1984fy,Kuzmin:1985mm},  b) anomalous baryon number violating processes in electroweak baryogenesis~\cite{Kuzmin:1985mm,Shaposhnikov:1987tw,Cohen:1993nk,Trodden:1998ym,Dine:2003ax}, and c) heavy right-handed Majorana neutrinos in leptogenesis~\cite{Fukugita:1986hr,Davidson:2008bu}. Developing novel baryogenesis mechanisms has been one of the main focuses of particle physics in the past decades.

The SM also needs an extension to explain the smallness of CP violation in QCD~\cite{Baker:2006ts} which on theoretical grounds is expected to be large~\cite{tHooft:1976rip}. This is known as the strong CP problem and can be elegantly solved by the Peccei-Quinn (PQ) mechanism~\cite{Peccei:1977hh,Peccei:1977ur}. The so-called PQ symmetry is spontaneously broken to yield a pseudo Nambu-Goldstone boson, the axion~\cite{Weinberg:1977ma,Wilczek:1977pj}. The PQ symmetry is explicitly broken by the quantum effects of QCD of the Adler-Bell-Jackiw type~\cite{Adler:1969gk,Bell:1969ts}. The quantum effects give a potential to the axion and drive the axion field value to the point where CP symmetry is restored, solving the strong CP problem. The PQ mechanism is especially attractive because the axion is also a dark matter candidate~\cite{Preskill:1982cy,Abbott:1982af,Dine:1982ah}, which provides yet another missing piece of the Standard Model.

We discover that when the PQ mechanism is introduced into the SM, the baryon ($B$) and lepton ($L$) asymmetries are generated in a wide class of models.
We call the following baryogenesis scheme as axiogenesis, which in general includes two main ingredients: 1) an asymmetry of the PQ charge is generated in the early universe as a coherent rotation in the axion direction and 2) the PQ asymmetry is later transferred to the $B+L$ asymmetry via the QCD and electroweak sphaleron transitions.
(We may convert the $B+L$ asymmetry into the $B-L$ asymmetry by some $B-L$ breaking interaction. Such a scenario will be investigated in a future work~\cite{Co:axio}.)
We contrast axiogenesis with other existing baryogenesis models after we introduce a concrete example.

The PQ symmetry is an approximate global symmetry which is explicitly broken by the QCD anomaly. Given that the symmetry is not exact, it is plausible that the PQ symmetry is significantly broken in the early universe, and the rotation of the axion is induced. In fact, it is expected that quantum gravity does not allow for a global symmetry~\cite{Giddings:1988cx,Coleman:1988tj,Gilbert:1989nq,Harlow:2018jwu,Harlow:2018tng} and the PQ symmetry is at best understood as an accidental symmetry explicitly broken by higher dimensional operators~\cite{Holman:1992us,Barr:1992qq,Kamionkowski:1992mf,Dine:1992vx}. Even when one requires that this explicit breaking not spoil the solution to the strong CP problem in the present universe, the rotation can still be induced from such interactions in the early universe as we will describe. Another example is a larger QCD scale in the early universe~\cite{Dvali:1995ce,Banks:1996ea,Co:2018phi,Co:2018mho}, which can initiate the axion oscillation and, once the QCD scale becomes small enough, the axion begins to rotate. These PQ-breaking sources well justify the axion rotation.

A fast rotation of the axion corresponds to a large PQ charge asymmetry.
The PQ symmetry and the SM quark chiral symmetries are explicitly broken by the quantum effect of QCD, called the QCD anomaly.
In the thermal bath of the early universe, a non-perturbative process called the QCD sphaleron transition is active. The transition, through the anomaly, converts the PQ charge asymmetry into the quark chiral asymmetry until the asymmetries reach equilibrium values.
The quark chiral symmetry and the $B+L$ symmetry are also explicitly broken by a weak anomaly. 
The quark chiral asymmetry is then converted into the $B+L$ asymmetry by another non-perturbative process known as the electroweak sphaleron transition. We may also consider a model with a weak anomaly of the PQ symmetry, as is the case with the KSVZ model~\cite{Kim:1979if,Shifman:1979if} embedded into grand unification and the supersymmetric DFSZ model~\cite{Zhitnitsky:1980tq,Dine:1981rt}.  In such a model the PQ asymmetry is directly converted into the $B+L$ asymmetry via electroweak sphaleron transitions. 
Consequently, the rotation of the axion can account for the observed matter asymmetry of the universe via the QCD and electroweak sphaleron transitions.

{\bf Baryon asymmetry from axion rotation.}---%
We discuss a minimal version of axiogenesis that achieves
the conversion between the PQ asymmetry $n_{\rm PQ}$ in the form of the axion rotation $\dot{\theta}$ and the baryon asymmetry solely by the SM QCD and electroweak sphaleron processes.

The axion $\phi_a$ is the angular direction of the complex scalar field 
\begin{equation}
P = \frac{1}{\sqrt{2}} \left( S+f_a \right) e^{i \frac{\phi_a}{f_a}}
\end{equation}
whose radial direction obtains a vacuum expectation value $f_a$, which is called the axion decay constant, and breaks the PQ symmetry. Analogous to how classical rotational symmetry leads to angular momentum conservation, the shift symmetry $\phi_a \rightarrow \phi_a + \alpha f_a$ implies a conserved Noether charge associated with the rotation in the axion direction. The PQ charge asymmetry $n_{\rm PQ}$ is exactly the Noether charge density associated with the shift symmetry. We define $n_{\rm PQ}$ with the following normalization, $n_{\rm PQ} =  i P \dot{P}^*-i P^* \dot{P}$, where the dot denotes a time derivative. When the radial mode is settled to the minimum $f_a$, the PQ charge asymmetry is then given by
\begin{equation}
\label{eq:nPQ}
n_{\rm PQ} =  \dot{\theta} f_a^2 ,
\end{equation}
where $\theta \equiv \phi_a / f_a$. Here we simply assume the rotation exists, while we present a concrete model to initiate the axion rotation in the next section.

The PQ asymmetry is converted into chiral asymmetries of SM quarks via QCD sphaleron transitions. The chiral asymmetries are then converted into the $B+L$ asymmetry via electroweak sphaleron transitions. Although the chiral symmetries are explicitly broken by the SM Yukawa couplings and hence the asymmetries are constantly washed out, the large PQ asymmetry continuously sources the chiral asymmetries and a nonzero baryon asymmetry remains in a quasi-equilibrium state.
If the PQ symmetry has a weak anomaly, the PQ asymmetry is directly converted into $B+L$ asymmetry.
In short, the PQ asymmetry is converted into $B+L$ asymmetry by QCD and electroweak sphaleron transitions.
With the detail given in the Supplemental Material, we find that, before the electroweak phase transition, the baryon number density $n_B$ is given by 
\begin{equation}
n_B = c_B \dot{\theta}T^2, \hspace{0.4 in} c_B \simeq 0.1 - 0.15 \, c_W.
\end{equation}
Here $c_W$ is the weak anomaly coefficient of the PQ symmetry normalized to that of the QCD anomaly.
The electroweak sphaleron process becomes ineffective after the electroweak phase transition and the baryon asymmetry is frozen. The resultant asymmetry normalized by the entropy density $s$ is
\begin{equation}
\label{eq:YB_thetadot}
Y_B = \frac{n_B}{s} = \left. \frac{45 c_B}{2 g_* \pi^2} \frac{\dot{\theta}}{T} \right|_{T = T_{\rm ws}} \hspace{-0.2 in} \simeq 2\times 10^{-3} \left( \frac{c_B}{0.1} \right) \frac{\dot{\theta}(T_{\rm ws})}{T_{\rm ws}} ,
\end{equation}
where $T_{\rm ws}$ is the temperature below which the electroweak sphaleron transition becomes ineffective and $g_*$ is the effective degrees of freedom in the thermal bath.

For $\dot{\theta}$ required to reproduce the baryon asymmetry, the axion continues to rapidly rotate even around the QCD phase transition.
Even when the axion mass becomes comparable to the Hubble expansion rate, the oscillation does not occur because the kinetic energy of the rotation is still much larger than the barrier of the axion cosine potential. The actual oscillation around the minimum is delayed until when the kinetic energy becomes comparable to the potential energy of the axion. Therefore, the axion abundance becomes enhanced~\cite{Co:2019jts} in comparison with the conventional misalignment mechanism~\cite{Preskill:1982cy,Abbott:1982af,Dine:1982ah}.

As derived in the Supplemental Material, assuming PQ charge conservation, $\dot{\theta}$ is a constant before the PQ breaking field reaches the minimum, whereas $\dot{\theta} \propto a^{-3}$ thereafter, with $a$ the scale factor. Assuming the latter case at the weak scale, we find the axion abundance
\begin{equation}
\label{eq:axion_DM}
\frac{\Omega_a h^2}{\Omega_{\rm DM} h^2} \simeq 140 \ \left( \frac{f_a}{10^8~{\rm GeV}} \right) \left( \frac{130~{\rm GeV}}{T_{\rm ws}} \right)^2 \left( \frac{0.1}{c_B} \right),
\end{equation}
to be much larger than the observed DM abundance $\Omega_{\rm DM} h^2$ for $f_a$ satisfying the astrophysical constraints~\cite{Ellis:1987pk,Raffelt:1987yt,Turner:1987by,Mayle:1987as,Raffelt:2006cw,Payez:2014xsa,Bar:2019ifz},
the SM prediction $T_{\rm ws} \simeq 130$ GeV \cite{DOnofrio:2014rug}, and $c_B = \mathcal{O}(0.1\mathchar`-1)$. (A value of $f_a = \mathcal{O} (10^6) \GeV$ leads to both successful axiogenesis and axion dark matter and interestingly resides in the so-called ``axion hadronic window''~\cite{Turner:1987by, Engel:1990zd, Chang:1993gm}, which however is recently under scrutiny~\cite{Chang:2018rso, Carenza:2019pxu}.) We require either 1) the axion rotation is damped after the electroweak phase transition, 2) the electroweak phase transition occurs earlier than the SM prediction, or 3) $c_B \gg \mathcal{O}(1)$ because of a large coefficient of the weak anomaly.

When the Higgs couples to particles with masses above the electroweak scale, it is possible that the electroweak phase transition occurs at a high temperature, and the Higgs eventually relaxes to the electroweak scale. We present a toy model in the Supplemental Material.

A large weak anomaly coefficient is possible in multi-field extensions of the Kim-Nilles-Peloso mechanism~\cite{Kim:2004rp,Harigaya:2014eta,Choi:2014rja,Harigaya:2014rga,Choi:2015fiu,Kaplan:2015fuy}, as considered in~\cite{Farina:2016tgd}. Assuming axion dark matter, the axion-photon coupling is 
\begin{equation}
\left|g_{a\gamma \gamma} \right| = \frac{ \alpha \left( c_W + c_Y \right)}{2 \pi f_a} \simeq 10^{-9} \GeV^{-1} \left( \frac{130~{\rm GeV}}{T_{\rm ws}}  \right)^2
\end{equation}
where $\alpha$ is the fine structure constant.
This prediction assumes that the hypercharge anomaly coefficient $c_Y$ of the PQ symmetry is negligible. For $T_{\rm ws} = 130 \GeV$, this large coupling is excluded by the limit from CAST~\cite{Anastassopoulos:2017ftl}, $\left|g_{a\gamma \gamma} \right| < 6.6 \times 10^{-11} \GeV^{-1}$. However, the contribution from the hypercharge anomaly can reduce or even exactly cancel the coupling.

We treat the rotation as a background field. A small portion of the PQ asymmetry is converted into the quark chiral asymmetries which are washed out by the Yukawa couplings.
The washout interaction is suppressed by a small up quark Yukawa coupling $y_u$ because in the limit of a vanishing $y_u$, a linear combination of the PQ symmetry and the up quark chiral symmetry is exact and washout does not occur. As is shown in the Supplemental Material, the washout of the PQ asymmetry is negligible.

We comment on the similarities and the differences of axiogenesis with the models in the literature.
In spontaneous baryogenesis~\cite{Cohen:1987vi,Cohen:1988kt}, baryon asymmetry is generated by a chemical potential of baryons given by the motion of a pseudo Nambu-Goldstone boson. The chemical potential is provided by the oscillation or the slow motion of the boson field driven by an explicit symmetry breaking potential. In axiogenesis, explicit breaking is effective only at higher energy scales and drives the rapid rotation of the axion instead. As a result, axiogenesis is compatible with the QCD axion.
Also, in spontaneous baryogenesis the oscillation itself washes out the PQ asymmetry, and the $B+L$ asymmetry needs to be converted into $B-L$ asymmetry e.g.~by the seesaw operator, which is not required in axiogenesis. Baryogenesis using the chemical potential provided by the rotation of the QCD axion is mentioned in~\cite{Takahashi:2003db} but the conversion of the PQ asymmetry into the $B+L$ asymmetry by the QCD and/or weak anomaly is not considered. Baryogenesis via the oscillation of the (QCD) axion by a large mass, the weak anomaly of the PQ symmetry and the seesaw operator is considered in~\cite{Kusenko:2014uta}. Baryogenesis by the chemical potential of the weak Chern-Simon number is utilized in the local electroweak baryogenesis~\cite{Turok:1990in,McLerran:1990zh} and other models in~\cite{Kuzmin:1992up,Servant:2014bla,Ipek:2018lhm}, where the chemical potentials are provided by the Higgs fields and the gluon condensation, respectively.

{\bf Affleck-Dine Axiogenesis.}---%
In this section we continue the investigation of a concrete realization of axiogenesis by evaluating $\dot{\theta}$.
To generate the PQ asymmetry, we employ the idea of Affleck-Dine~\cite{Affleck:1984fy} proposed in a supersymmetric theory, even though supersymmetry is not essential to axiogenesis. (See~\cite{Harigaya:2019emn} for a non-supersymmetric Affleck-Dine mechanism.)
For clarity and simplicity, we demonstrate a working example by the quartic potential
\begin{equation}
\label{eq:quartic}
V =  \lambda^2 \left( |P|^2 - \frac{f_a^2}{2} \right)^2, \hspace{0.3 in} \lambda^2 = \frac{1}{2} \frac{m_S^2}{f_a^2},
\end{equation}
where $P$ is the complex field breaking the PQ symmetry in the vacuum and $m_S$ corresponds to be the vacuum mass of the radial mode $S$, which we call the saxion although we do not assume supersymmetry here. The angular mode in the vacuum is the axion. We assume a large initial field value $|P_i| = S_i / \sqrt{2} \gg f_a$, which can arise for a sufficiently small quartic coupling, namely a flat potential of $S$.
(A flat potential is natural in supersymmetric theories, with which we demonstrate axiogenesis using a concrete model and cosmological evolution in the Supplemental Material.)
The potential at large field values is dominated by the quartic term and thus the saxion mass is initially given by $\sqrt{3} \lambda S_i$. The saxion starts oscillating when the Hubble friction drops below the mass, $3 H \simeq \sqrt{3} \lambda S_i$, at the temperature 
\begin{equation}
\label{eq:Tosc_Si}
%T_{\rm osc} \simeq 2 \times 10^{12}~{\rm GeV} \left( \frac{S_i}{10^{17}~{\rm GeV}} \right)^{ \scalebox{1.01}{$\frac{1}{2}$} }  \left( \frac{\lambda }{10^{-10}} \right)^{ \scalebox{1.01}{$\frac{1}{2}$} }.
T_{\rm osc} = \left( \frac{30}{\pi^2 g_*} \right)^{ \scalebox{1.01}{$\frac{1}{4}$} }  \sqrt{\lambda M_{\rm Pl} S_i} ,
\end{equation}
with $M_{\rm Pl} = 2.4 \times 10^{18}$ GeV the reduced Planck constant.

For large $S_i$, a higher dimensional potential term that explicitly breaks the PQ symmetry,
\begin{align}
V_{\cancel{\rm PQ}}= \frac{P^n}{M^{n-4}} + {\rm h.c.},
\end{align}
can be effective. Here $M$ is a dimensionful constant.
%If $S_i$ is sufficiently close to the suppression scale of the higher dimensional potential term that breaks the PQ symmetry, the potential can be effective in driving $P$ in the angular direction and cause a rotation.
The potential drives $P$ in the angular direction and causes a rotation.
After initiating the rotation, as $S$ decreases by redshifting, explicit PQ breaking quickly becomes very suppressed as it originates from a higher dimensional operator. As a result, the PQ charge becomes conserved soon after the initial motion. It is convenient to normalize the asymmetry by the number density of the saxion,
\begin{equation}
\frac{n_{\rm PQ}}{n_S} \equiv \epsilon
\end{equation}
because this is a redshift-invariant quantity. The scaling of $n_{\rm PQ} \propto a^{-3}$ can be understood as a result of PQ charge conservation. We use $\epsilon$ to parametrize the amount of PQ breaking that leads to the axion rotation or equivalently the potential gradient in the angular direction relative to that of the radial mode at $S=S_i$. We treat $\epsilon$ as a free parameter in what follows. In supersymmetric theories described in the Supplemental Material, $\epsilon$ is naturally order unity.

The saxion acquires a large energy density due to its initial condition $S_i \gg f_a$. While the saxion condensate will eventually thermalize with the SM plasma, as is shown in the Supplemental Material, the PQ charge asymmetry is conserved up to cosmic expansion. In other words, thermalization only depletes the energy density in the radial mode and preserves that in the angular mode. Therefore, the rotation continues even after thermalization. Whether the saxion condensate dominates the energy density of the universe before being thermalized into the SM plasma leads to two possibilities for the subsequent cosmology, both of which we investigate in order. 

If the universe stays radiation-dominated throughout the evolution, the PQ asymmetry due to the axion rotation in units of the entropy density is a redshift-invariant quantity after the onset of the oscillation and is given by
\begin{equation}
\label{eq:n_theta}
%Y_{\rm PQ} = \frac{n_{\rm PQ}}{s}  = \epsilon \frac{V(P_i)}{s \, m_S (P_i)} = \epsilon \frac{ 15 \sqrt{3} }{ 8 g_* \pi^2 } \frac{\lambda S_i^3}{T_{\rm osc}^3 }.
Y_{\rm PQ}  = \frac{\epsilon V(P_i)}{s \, m_S (P_i)} = \epsilon \left( \frac{\pi^2 g_*}{30} \right)^{ \scalebox{1.01}{$\frac{3}{4}$} } \frac{ 15 \sqrt{3} }{ 8 \pi^2 g_* } \frac{ S_i^{3/2}}{\sqrt{\lambda} M_{\rm Pl}^{3/2} } ,
\end{equation}
which, with Eq.~(\ref{eq:nPQ}), implies that the angular speed is
\begin{equation}
\dot\theta (T) =  \epsilon \left( \frac{\pi^2 g_*}{30} \right)^{ \scalebox{1.01}{$\frac{3}{4}$} } \frac{ S_i^{3/2} T^3}{4 \sqrt{3} \sqrt{\lambda} M_{\rm Pl}^{3/2} \feff^2(T) },
\end{equation}
where $\feff (T)$ is the effective axion decay constant at temperature $T$, i.e.~$\sqrt{2} \left|P\left(T\right)\right|$.
Finally, based on Eq.~(\ref{eq:YB_thetadot}), the baryon asymmetry is evaluated at the temperature when the electroweak sphaleron is out of equilibrium and reads
\begin{align}
\label{eq:YB_RD}
 Y_B & =  \epsilon \left( \frac{\pi^2 g_*}{30} \right)^{ \scalebox{1.01}{$\frac{3}{4}$} }  \frac{ 15 \sqrt{3} }{ 8 \pi^2 g_* } \frac{ c_B S_i^{3/2} T_{\rm ws}^2}{\sqrt{\lambda} M_{\rm Pl}^{3/2} \feff^2(T_{\rm ws}) } \\
& \simeq 9 \times 10^{-11} \epsilon \, \xi \left( \frac{S_i}{M_{\rm Pl}} \right)^{ \scalebox{1.01}{$\frac{3}{2}$} } \left( \frac{10^9 \GeV}{ \feff(T_{\rm ws}) } \right)^{ \scalebox{1.01}{$\frac{3}{2}$} } \left( \frac{{\rm TeV}}{ m_S } \right)^{ \scalebox{1.01}{$\frac{1}{2}$} } , \nonumber
\end{align}
where
\begin{align}
\label{eq:xi}
\xi \equiv \left( \frac{c_B}{100} \right)  \left( \frac{T_{\rm ws} }{ 130 \GeV} \right)^2.
\end{align}
As illustrated in the previous section, the dark matter abundance in Eq.~(\ref{eq:axion_DM}) demands a transfer of the PQ to baryon asymmetries more efficient than that from the SM prediction of $T_{\rm ws} = 130$ GeV along with $c_B \simeq \mathcal{O}(0.1$-$1)$ from a weak anomaly coefficient $c_W$ of order unity. This is manifest in the parameter $\xi$, and theories with a large $c_B$ and/or $T_{\rm ws}$ are also discussed previously.

\begin{figure}
	\includegraphics[width= \linewidth]{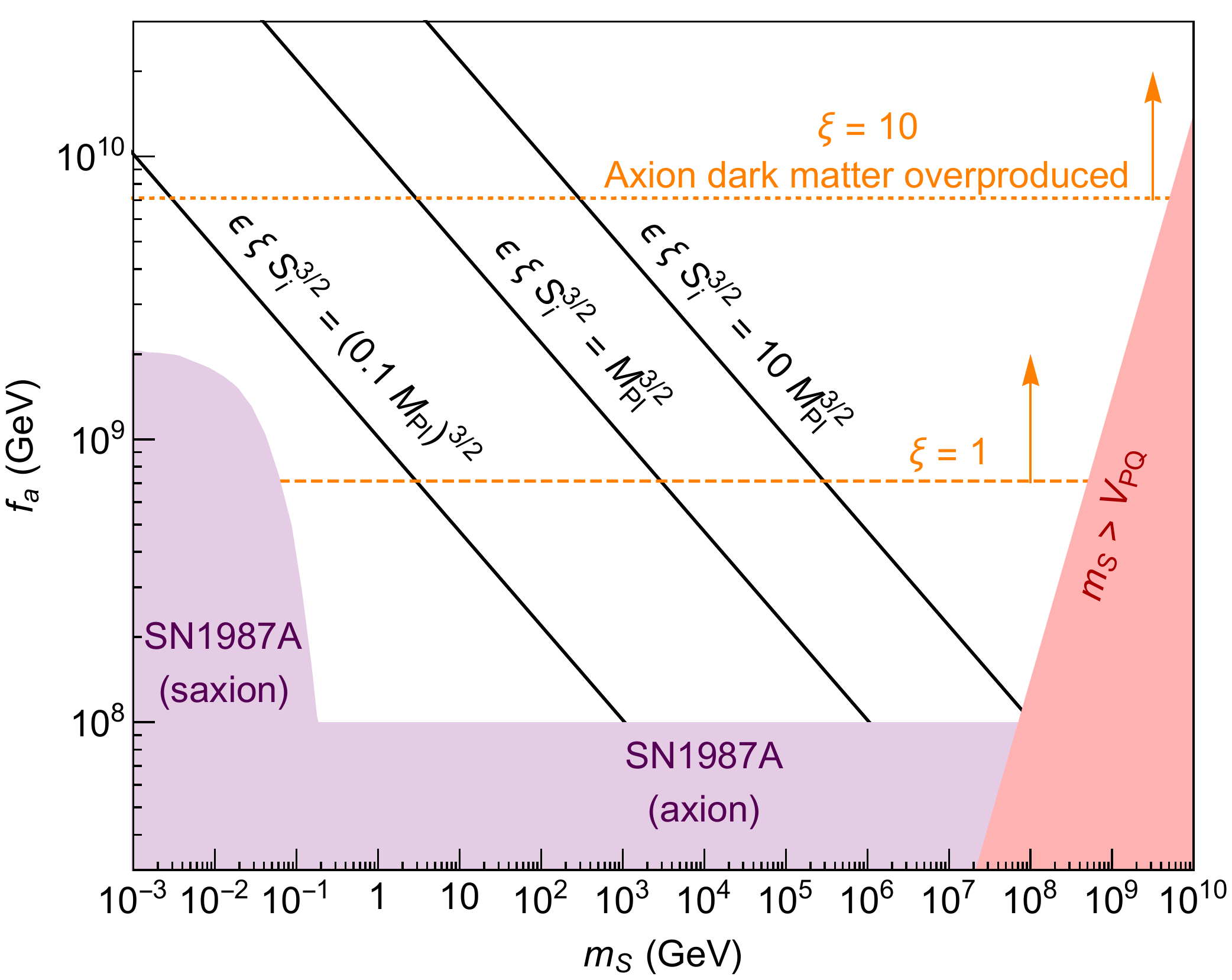}
	\caption{The parameter space compatible with the observed baryon asymmetry.}
	\label{fig:fa_ms}	
\end{figure}

In Fig.~\ref{fig:fa_ms}, we demonstrate the viable parameter space of the saxion mass $m_S$ and the axion decay constant $f_a$. The black contours show the values of $\epsilon \, \xi  S_i^{3/2}$ required by Eq.~(\ref{eq:YB_RD}) for the observed baryon asymmetry $Y_B^{\rm obs} = 8.7 \times 10^{-11}$~\cite{Aghanim:2018eyx}. From Eq.~(\ref{eq:axion_DM}), the region above the orange line is excluded due to axion dark matter overproduction for $\xi = 1$ (dashed) and $\xi = 10$ (dotted). In the red region, the saxion mass $m_S$ exceeds the unitarity limit.
The purple region is excluded since the emission of saxions or axions in a supernova core affects the duration of the neutrino emission~\cite{Ellis:1987pk,Raffelt:1987yt,Turner:1987by,Mayle:1987as,Ishizuka:1989ts,Raffelt:2006cw,Chang:2018rso,Carenza:2019pxu}. The constraint from the saxion emission can be, however, evaded by introducing a large enough saxion-Higgs mixing to trap saxions inside the core.

If the saxion dominates, since the $P$ oscillation until thermalization at temperature $T_{\rm th}$, the PQ charge number density $n_{\rm PQ}$ and the saxion number density $n_{\rm S}$ redshift the same way. After the saxion is depleted to create a thermal bath with a temperature $T_{\rm th}$, the yield of the PQ asymmetry remains a constant given by
\begin{equation}
Y_{\rm PQ} = \epsilon \frac{3 T_{\rm th}}{4 m_S} .
\end{equation}
Similarly, with Eq.~(\ref{eq:nPQ}), the angular speed is
\begin{equation}
\dot{\theta}(T) = \epsilon \frac{g_* \pi^2 }{30} \frac{T_{\rm th}T^3}{m_S \feff^2(T)}.
\end{equation}
Based on Eq.~(\ref{eq:YB_thetadot}), we obtain
\begin{align}
\label{eq:YB_MD}
Y_B & =  \epsilon  \frac{ 3 c_B T_{\rm th} T_{\rm ws}^2 }{4 m_S \feff^2(T_{\rm ws})} \\
& \simeq 10^{-10} \epsilon \, \xi \left( \frac{T_{\rm th}}{100 \GeV} \right) \left( \frac{10^9 \ GeV}{ \feff(T_{\rm ws}) } \right)^2 \left( \frac{{\rm TeV}}{ m_S } \right) . \nonumber
\end{align}
This expression is valid whether thermalization or the electroweak phase transition occurs first and is also general for any type of the potential. While contours of $\epsilon \, \xi \, T_{\rm th}$ can be easily included in Fig.~\ref{fig:fa_ms}, a concrete model is necessary to realize the required values of $T_{\rm th}$. In the Supplemental Material, we thoroughly demonstrate a consistent thermalization history required by Eq.~(\ref{eq:YB_MD}) for the observed baryon asymmetry in the framework of supersymmetry. Supersymmetry is again motivated by the flatness of the potential, or equivalently a light saxion, to obtain a large saxion initial field value. A large viable parameter space is similarly obtained in the supersymmetric version of axiogenesis. In summary, Fig.~\ref{fig:fa_ms} shows that a wide range of the saxion mass $m_S$ is viable, while a low $f_a$ is favored in the minimal realization of axiogenesis. 

{\bf Discussion.}---%
We propose a mechanism to explain the baryon asymmetry of the Universe. The two main ingredients are a rotation in the axion direction in the early Universe, corresponding to an excess of PQ charges, as well as QCD and electroweak sphaleron processes that convert the PQ asymmetry into those of baryons and leptons. We construct a concrete model where the rotation is a consequence of higher dimensional PQ-breaking operators. This is analogous to how the rotation of the Affleck-Dine field arises. We show that a sufficient baryon asymmetry is generated from the PQ charge by the QCD and electroweak sphaleron transitions.

Intrinsic to the axiogenesis framework, the angular speed of the rotation needed for the observed baryon asymmetry leads to axion dark matter. In fact, axion dark matter is overproduced in the minimal scenario where the weak anomaly coefficient of the PQ symmetry is as large as the QCD anomaly coefficient and the PQ charge is conserved even after the electroweak sphaleron transition becomes ineffective, which the Standard Model predicts to be at $T_{\rm ws} = 130 \GeV$. Therefore, unless the PQ charge is depleted after the electroweak phase transition, the associated prediction is a value of $T_{\rm ws}$ that is higher than predicted by the Standard Model and/or a large weak anomaly coefficient.
We show how new physics at the $1\mathchar`-10 $ TeV scale can raise $T_{\rm ws}$ so that the axion can constitute a subdominant or correct amount of dark matter. In addition to new heavy states, axiogenesis also favors a small decay constant which is accessible to many axion haloscope and helioscope experiments~\cite{Vogel:2013bta, Armengaud:2014gea, Arvanitaki:2014dfa, Rybka:2014cya, Sikivie:2014lha, TheMADMAXWorkingGroup:2016hpc, McAllister:2017lkb, Anastassopoulos:2017kag, Arvanitaki:2017nhi, Geraci:2017bmq, Baryakhtar:2018doz,Du:2018uak, Marsh:2018dlj}. The evolution of the PQ breaking field reveals non-standard cosmological eras, which alone may have profound implications for other aspects of cosmology. These phenomenological prospects render axiogenesis an exciting avenue to pursue theoretically and experimentally.

{\bf Acknowledgment.}---%
We are grateful to Nicolas Fernandez and Akshay Ghalsasi for collaboration in the early stages of this work. We thank Lawrence J. Hall and Aaron Pierce for useful discussions and comments on the manuscript. The work was supported in part by the DoE Early Career Grant DE-SC0019225 (R.C.), the DoE grant DE-SC0009988 (K.H.) and the Raymond and Beverly Sackler Foundation Fund (K.H.).

\bibliography{axiogenesis}

\clearpage
\newpage
\maketitle
\onecolumngrid
\begin{center}
\textbf{\large Axiogenesis} \\ 
\vspace{0.05in}
{ \it \large Supplemental Material}\\ 
\vspace{0.05in}
{Raymond T. Co and Keisuke Harigaya}
\end{center}
\onecolumngrid
%%%%%%%%%% Merge with Supplemental material %%%%%%%%%%
\setcounter{equation}{0}
\setcounter{figure}{0}
\setcounter{table}{0}
\setcounter{section}{0}
\setcounter{page}{1}
\makeatletter
\renewcommand{\theequation}{S\arabic{equation}}
\renewcommand{\thefigure}{S\arabic{figure}}

This Supplemental Material is organized as follows. In Secs.~\ref{sec:therm_rot}, \ref{sec:washout_ss}, and \ref{sec:evol_rot}, we analyze the evolution of the rotating Peccei-Quinn symmetry breaking field. The discussions in  Secs.~\ref{sec:therm_rot} and \ref{sec:evol_rot} are applicable to generic $U(1)$ symmetric theories. In Sec.~\ref{sec:baryon_asymm}, we derive the formula for the baryon asymmetry from the rotating Peccei-Quinn symmetry breaking field. In Sec.~\ref{sec:EWPT}, we present a model where the electroweak phase transition occurs at a high temperature. In Sec.~\ref{sec:SUSY_axio}, we construct a supersymmetric realization of axiogenesis and discuss the details of the cosmological evolution.

%%%%%%%%%%%%%%%%%%%%%%%%%%%
\section{Thermodynamics of a rotating field}
\label{sec:therm_rot}
%%%%%%%%%%%%%%%%%%%%%%%%%%%%
We will show by thermodynamics that the $U(1)$ charge stored in the rotation of such a $U(1)$ symmetry breaking field will retain most of the charge even when it is in thermal equilibrium via some efficient particle interactions that conserve the $U(1)$ charge.

Suppose that a complex scalar $P$ charged under $U(1)$ symmetry is initially rotating, i.e.~$P=V_0 e^{ i \dot{\theta}_0t}$, and that $P$ is coupled to and in equilibrium with a thermal bath at temperature $T$. We assume $V_0 \gg T$. To be concrete, we consider a case where an interaction transfers a unit charge of $P$ into charges for Weyl fermions $\psi_i$ ($i = 1, \dots, N$) in a charge-conserving manner,
\begin{align}
\label{eq:transfer}
P \leftrightarrow \psi_1\psi_2 \cdots \psi_N.
\end{align}
The following discussion is applicable to generic interactions which transfer the charge of $P$ into the asymmetries of scalar excitations or multiple identical particles.
We for now assume that the asymmetry of $\psi_i$ is conserved up to the above interaction and consider a more generic setup in the next section.

We derive the asymmetry assuming thermal equilibrium where the charge of $P$ is partially converted into the asymmetry of $\psi_i$, defined as the difference in the number densities of particles and antiparticles.
We parametrize the initial total $U(1)$ charge density by $2V_0^2 \dot{\theta}_0\equiv n_0$. We will denote particle $\psi_i$'s equilibrated asymmetry by $n_{\psi_i} = n_{\psi}$ and the rotating field's charge density by $n_0 -  n_{\psi} $. We denote the chemical potential of $\psi_i$ by 
\begin{equation}
\mu_{\psi_i}  = 6 \frac{n_{\psi}}{T^2},
\end{equation}
which we assumed to be much smaller than $T$.
We now solve for $n_\psi$ by minimizing the free energy. The energy, pressure, and entropy of the thermal bath are
\begin{align}
\rho = \frac{\pi^2}{30} g_* T^4 + 9 N \frac{n_{\psi}^2}{T^2}, \hspace{0.4 in} p  = \frac{\rho}{3}, \hspace{0.4 in} s = \frac{\rho +  p - \sum\limits_{i} \mu_{\psi_i} n_{\psi_i}}{T} = \frac{2\pi^2}{45} g_* T^3 + 6 N \frac{n_{\psi}^2}{T^3} .
\end{align}
The free energy of the thermal bath ${F_{\rm th}}$ per volume $\mathcal{V}$ is
\begin{align}
\frac{F_{\rm th}}{\mathcal{V}} = \rho - T s  = - \frac{\pi^2}{90} g_* T^4 + 3 N \frac{n_{\psi}^2}{T^2} .
\end{align}
For $V_0 = V_{\rm vac}$, where $V_{\rm vac}$ is the vacuum expectation value of $P$, $\dot{\theta} = (n_0 - n_\psi )/ (2 V_{\rm vac}^2)$ and the energy density of the rotation is
\begin{align}
\rho_{\rm rot} = V_{\rm vac}^2 \dot{\theta}^2 = \frac{1}{4V_{\rm vac}^2} \left(n_0 - n_\psi \right)^2.
\end{align}
For $V_0 \gg V_{\rm vac}$, the rotational speed is given by the curvature of the potential $m$, i.e.~$\dot{\theta} = m$, so the energy density of the rotation is given by 
\begin{align}
\rho_{\rm rot} = m \left( n_0 - n_\psi \right) .
\end{align}
The entropy of the rotation is zero, and hence the free energy density of the system is given by $F_{\rm th}/\mathcal{V} + \rho_{\rm rot}$. For both $V_0 = V_{\rm vac}$ and $V_0 \gg V_{\rm vac}$, the free energy is minimized for
\begin{align}
n_{\psi } = \frac{1}{12 N }\frac{T^2}{V_0^2}n_0 = \frac{1}{6 N} \dot{\theta} T^2 .
\end{align}
Note that $n_\psi \ll n_0$, and hence the rotating field cannot lose a significant fraction of its charge even when coupled to the thermal bath.

The equilibrium value of $n_\psi$ and the detailed balance mean that the contribution of the interaction in Eq.~(\ref{eq:transfer}) to the Boltzmann equation is
\begin{align}
\dot{n}_{\psi_i} = - \Gamma \left( \sum_in_{\psi_i} - \frac{1}{6}\dot{\theta}  T^2 \right),
\end{align}
where $\Gamma$ is the rate of the interaction. When there are other interactions that transfer charges in addition to Eq.~(\ref{eq:transfer}) and/or the equilibrium is not reached, one can solve the Boltzmann equation including this term to find the equilibrium value or the evolution of the charge asymmetry.

When the fermions $\psi_i$ are charged under gauge symmetries, the asymmetries of the fermions are transferred into magnetic helicity of gauge fields~\cite{Joyce:1997uy}. In axiogenesis, the chiral asymmetries of fermions are of the order of the baryon asymmetry. The transfer rate is much smaller than the Hubble expansion rate~\cite{Long:2016uez} in our setup because the chemical potential is much lower than the temperature and we neglect such a transfer in this work.

We have implicitly assumed that the temperature of the thermal bath remains the same in order to find the equilibrium state by minimizing the free energy. For $n_0 \gtrsim T V_0^2$, this assumption breaks down and, simultaneously, the approximation $\mu_\psi/T \ll 1$ also breaks down. It is necessary to treat the whole system as an isolated system and maximize entropy to find the equilibrium state. We emphasize that most of the $U(1)$ asymmetry remains in the rotation. The thermal bath with asymmetry $n_\psi$ has an energy density larger than $n_\psi^{4/3}$. For $n_0 > T V_0^2$, the initial energy is dominated by the rotation, and $n_\psi^{4/3} < n_0^2/ V_0^2 $, which means $n_\psi < n_0 (n_0/ V_0^3)^{1/2} < n_0$, where we assume that the mass of the $U(1)$ symmetry breaking field is below $V_0$.

%%%%%%%%%%%%%%%%%%%%%%%%%%%
\section{Washout effects from strong sphaleron processes}
\label{sec:washout_ss}
%%%%%%%%%%%%%%%%%%%%%%%%%%%%
It is established in the previous section that the $U(1)$ charge in the form of a rotating field does not get depleted due to charge-conserving interactions. In this section, we extend the analysis to include possible washout effects due to quark Yukawa couplings and show that such washout effects are ineffective.

We first consider the coupling of the axion to gluons. We build up our intuition with only an up quark in the thermal bath, after which we generalize the derivation to include other Standard Model and heavy quarks charged under PQ symmetry.
The QCD sphaleron transition produces the chiral asymmetry of up quarks from the PQ asymmetry, and the chiral symmetry is washed out by the scattering involving the Yukawa coupling of the up quark. The Boltzmann equations for the PQ asymmetry $n_{PQ}$ and the chiral asymmetry of up quarks $n_u$ are
\begin{align}
\dot{n}_{\rm PQ} = - \Gamma_{\rm ss} \left( \frac{\dot{\theta}}{T} - \frac{n_u}{T^3}  \right) = -\Gamma_{\rm ss} \left( \frac{n_{\rm PQ}}{\veff^2 T} - \frac{n_u}{T^3}  \right), \hspace{0.5in}
\dot{n}_u = + \Gamma_{\rm ss} \left( \frac{n_{\rm PQ}}{\veff^2 T} - \frac{n_u}{T^3}  \right) - \alpha_3 y_u^2 T n_u,
\end{align}
where $\Gamma_{\rm ss}\sim \alpha_3^4 T^4$ is the QCD strong sphaleron transition rate per volume, and $\veff$ is the field value of the PQ symmetry breaking field, which in the early universe does not necessarily coincide with $\vpq$. We have omitted unimportant $\mathcal{O}(1)$ factors. Since the sphaleron transition is effective, $n_u$ quickly reaches the equilibrium value. By taking $\dot{n}_u=0$ and inserting the solution of $n_u$ to the Boltzmann equation of $n_{\rm PQ}$, we obtain
\begin{align}
\dot{n}_{\rm PQ} = - \Gamma_{\rm PQ} \, n_{\rm PQ}, \hspace{0.5 in} \Gamma_{\rm PQ} \simeq \alpha_3 \frac{y_u^2 T^3}{\veff^2}.
\end{align}
In the limit where the up quark Yukawa coupling vanishes, a linear combination of the PQ symmetry and the up quark chiral symmetry is exact and washout does not occur.
When there are several quarks $q$, with a similar computation where each $\dot{n}_q$ vanishes as well, one can show that 
\begin{align}
\label{eq:ss_washout}
\Gamma_{\rm PQ} \simeq \alpha_3 \frac{\tilde{y}^2 T^3}{\veff^2}, \hspace{0.5 in} \frac{1}{\tilde{y}^2} \equiv \sum_q \frac{1}{y_q^2} \simeq \frac{1}{y_u^2}.
\end{align}
The rate is determined by the smallest quark Yukawa coupling, namely, that of the up quark.
The washout rate is faster than the Hubble expansion rate during a radiation-dominated era if
\begin{align} 
T > 10^{12}~{\rm GeV } \left( \frac{\veff}{10^9 ~{\rm GeV}} \right)^2.
\end{align}
One may worry that washout is effective at high temperature. Fortunately, at such a high temperature, washout is ineffective because the PQ symmetry breaking field still has a large field value $\veff \gg \vpq$ as a result of the large initial field value set by the inflationary dynamics discussed in the Letter.

We next consider the case where the heavy quarks $Q\bar{Q}$ to which the PQ symmetry breaking field couples are in the thermal bath. As we will demonstrate, the presence of $Q\bar{Q}$ does not change the conclusion that washout of the PQ asymmetry is inefficient. We consider the coupling
\begin{align}
{\cal L} \supset - m_Q \frac{\veff}{\vpq} e^{- i \theta}Q\bar{Q}.
\end{align}
After performing the chiral rotations $Q\rightarrow e^{-i\theta/2}Q$ and $\bar{Q}\rightarrow e^{-i\theta/2}\bar{Q}$, the mass term does not depend on $\theta$, and instead the following couplings are induced,
\begin{align}
{\cal L} \supset - \frac{1}{2} \partial_\mu \theta \left( Q^\dag \bar{\sigma}^\mu Q + \bar{Q}^\dag \bar{\sigma}^\mu \bar{Q}  \right) + \frac{\theta }{64\pi^2} \epsilon^{\mu \nu \rho \sigma}G^a_{\mu\nu} G^a_{\rho \sigma}.
\end{align}

Let us for now ignore the Standard Model quarks. The chiral asymmetry of $Q\bar{Q}$, $n_Q$, and $n_{\rm PQ}$ evolve due to the QCD sphaleron transitions and the chirality flipping scattering by the mass term $m_Q$. The former conserves $2n_{\rm PQ} + n_Q$, while the latter conserves $2n_{\rm PQ} - n_Q$. Then the Boltzmann equations for $n_{\rm PQ}$ and $n_Q$ are
\begin{align}
\dot{n}_{\rm PQ} =& -\Gamma_{\rm ss} \left( \frac{n_{\rm PQ}}{\veff^2 T} - \frac{n_Q}{T^3} \right) - \alpha_3 m_Q^2 T^2 \left( \frac{n_{\rm PQ}}{\veff^2 T} + \frac{n_Q}{T^3} \right), \\
\dot{n}_{\rm Q} = & +2\Gamma_{\rm ss} \left( \frac{n_{\rm PQ}}{\veff^2 T} - \frac{n_Q}{T^3} \right) - 2 \alpha_3 m_Q^2 T^2 \left( \frac{n_{\rm PQ}}{\veff^2 T} + \frac{n_Q}{T^3} \right), \nonumber
\end{align}
where we omit unimportant $\mathcal{O}(1)$ factors. The factors of $2$ are important for the conservation laws and thus included. 

We now generalize the derivation to the following system of equations including both the up quark and heavy quarks
\begin{align}
\dot{n}_{\rm PQ} =& -\Gamma_{\rm ss} \left( \frac{n_{\rm PQ}}{\veff^2 T} - \frac{n_Q}{T^3} - \frac{n_u}{T^3} \right) - \alpha_3 m_Q^2 T^2 \left( \frac{n_{\rm PQ}}{\veff^2 T} + \frac{n_Q}{T^3} \right), \nonumber\\
\dot{n}_{\rm Q} = & +2\Gamma_{\rm ss} \left( \frac{n_{\rm PQ}}{\veff^2 T} - \frac{n_Q}{T^3} - \frac{n_u}{T^3} \right) - 2 \alpha_3 m_Q^2 T^2 \left( \frac{n_{\rm PQ}}{\veff^2 T} + \frac{n_Q}{T^3} \right), \\
\dot{n}_{\rm u} = & +2\Gamma_{\rm ss} \left( \frac{n_{\rm PQ}}{\veff^2 T} - \frac{n_Q}{T^3} - \frac{n_u}{T^3} \right) - \alpha_3 y_u^2 T n_u. \nonumber
\end{align}
Again, since the sphaleron transition is efficient, $n_Q$ and $n_u$ quickly reach equilibrium. By taking $\dot{n}_Q = \dot{n}_u=0$ and inserting the solution to the Boltzmann equation of $n_{\rm PQ}$, we obtain
\begin{align}
\dot{n}_{\rm PQ} \simeq - \Gamma_{\rm PQ} \,  n_{\rm PQ}, \hspace{0.5 in} \Gamma_{\rm PQ} = \frac{\alpha_3 y_u^2 T^3}{\veff^2} \frac{\alpha_3^3 m_Q^2}{\alpha_3^3 m_Q^2 + y_u^2 m_Q^2 + \alpha_3^3 y_u^2 T^2 } \simeq \frac{\alpha_3 y_u^2 T^3}{\veff^2} \frac{ m_Q^2}{m_Q^2 +  y_u^2 T^2 }.
\end{align}
The rate is no larger than that in Eq.~(\ref{eq:ss_washout}), the case without $Q\bar{Q}$ in the thermal bath. Therefore, the earlier conclusion that the PQ charge in the rotation is not depleted by thermal processes still applies even in the presence of the washout effects.

%%%%%%%%%%%%%%%%%%%%%%%%%%%
\section{Evolution of the energy density of the rotating field}
\label{sec:evol_rot}
%%%%%%%%%%%%%%%%%%%%%%%%%%%%
The PQ symmetry breaking field $P$ initially follows an elliptical orbit, meaning that both radial and angular motions are excited, and eventually thermalizes.
As we have shown in the previous sections, even in thermal equilibrium, most of the PQ charge is still stored in the rotation.
Thus, the thermalization of $P$ only partially depletes the energy density of $P$.
Then the trajectory of $P$ transitions from the elliptical one to the one that minimizes the energy for a fixed charge, namely a circular motion with a vanishing ellipticity.
In this section we show how the energy density of the circularly rotating PQ symmetry breaking field evolves in various eras.

After the circular motion is established by thermalization, as long as the frequency of the rotation is much larger than the Hubble expansion rate, the energy density of the PQ symmetry breaking field redshifts while the motion remains circular. This is essentially because there is no special direction in the complex plane of the PQ symmetry breaking field to which a major/minor axis of an elliptic motion can point to.

We denote the potential of the PQ symmetry breaking field as $V = f(|P|^2)$. The circular motion satisfying the equation of motion, neglecting the Hubble expansion, is
\begin{align}
P = \veff \, e^{ i \omega t}, \hspace{0.5 in} \omega ^2= f'(\veff^2),
\end{align}
where prime denotes $f'(|P|^2) \equiv {\rm d}(f(|P|^2))/{\rm d}(|P|^2)$ We take $\omega >0$ without loss of generality.
The kinetic energy density $K$, the total energy density $\rho$, and the number density $n_{\rm PQ}$ are
\begin{align}
K = |\dot{P}|^2 = \omega^2 \veff^2 =  f' \veff^2,  \hspace{0.5 in} 
\rho = f + f' \veff^2 ,  \hspace{0.5 in} 
n_{\rm PQ} = i P \dot{P^*} - i P^* \dot{P} = 2 \omega \veff^2  = 2 f'^{1/2} \veff^2.
\end{align}

Due to Hubble friction, $\veff$ changes slowly in comparison with the frequency of the circular motion. The number density decreases in proportion to $a^{-3}$,
\begin{align}
n_{\rm PQ} = 2 f'^{1/2} \veff^2 = n_0 \left( \frac{a_0}{a} \right)^3,
\end{align}
where $n_0$ is the number density when the scale factor $a = a_0$. Using this equation, we can use $a$ or $\veff$ as a time variable to describe the evolution of the rotation. By taking the derivative with respect to $\veff$ on both sides, we find
\begin{align}
a \frac{{\rm d} \veff}{{\rm d} a} = - \frac{3 \veff f'}{2 f' + \veff^2 f''}.
\end{align}

The dependence of the total energy density on $\veff$ is
\begin{align}
\frac{{\rm d} \rho}{{\rm d} \veff} = 2 \veff (2 f' + \veff^2 f''),
\end{align}
and the dependence on $a$ is
\begin{align}
a \frac{{\rm d} \rho}{{\rm d} a} = - 6 \veff^2 f' = - 6 K.
\end{align}
Note that this is consistent with the full equation of motion,
\begin{align}
\ddot{P} + 3 H \dot{P} + \frac{\partial V}{ \partial P^*} = 0.
\end{align}
The redshift scaling law derived from the equation of motion is
\begin{align}
\dot{\rho} = &   \ddot{P} \dot{P}^* + \ddot{P}^* \dot{P} + \frac{\partial V}{\partial P} \dot{P} + \frac{\partial V}{\partial P^*} \dot{P}^* = - 6  H |\dot{P}|^2 = - 6 H K,\nonumber \\
a \frac{{\rm d} \rho}{{\rm d} a}  = & \frac{\dot{\rho}}{H }  = - 6 K.
\end{align}

As an example, we consider the potential with the PQ symmetry broken by dynamical transmutation,
\begin{align}
V = m^2 |P|^2 \left( {\rm ln}\frac{|P|^2}{\vpq^2} -1  \right) + m^2\vpq^2.
\end{align}
Using the above equations, we find
\begin{align}
\frac{{\rm d} \ln \rho}{{\rm d} \ln a} = - 6 \frac{  \ln r^2 }{ 2  \ln r^2 - 1 + 1/ r^2} \rightarrow
\begin{cases}
 -3  & r \gg 1\\
  -6 & r\simeq 1
  \end{cases},
 \hspace{0.5 in} r \equiv \frac{\veff}{\vpq}.
\end{align}
When the PQ symmetry breaking field is rotating with a large radius $\veff \gg \vpq$, the energy density of the rotation redshifts as matter. As the radius approaches the vacuum expectation value $\vpq$, the rotation begins to behave as kination \cite{Spokoiny:1993kt, Joyce:1996cp}, and redshifts faster than radiation. Even if the energy density of the rotation dominates, the Universe eventually becomes radiation-dominated thanks to the fast redshift scaling of kination. In the intermediate stage, the Universe is kination-dominated.

%%%%%%%%%%%%%%%%%%%%%%%%%%%
\section{Baryon asymmetry}
\label{sec:baryon_asymm}
%%%%%%%%%%%%%%%%%%%%%%%%%%%%
In this section we explicitly show that nonzero baryon asymmetry is  created from the rotation of the PQ symmetry breaking field. For simplicity we only consider one generation of a quark doublet $q$, a right-handed up quark $\bar{u}$, a right-handed down quark $\bar{d}$, a lepton doublet $\ell$, a right-handed electron $\bar{e}$, and the Standard Model Higgs. The Yukawa interactions are
\begin{align}
{\cal L} = y_u H^\dag q \bar{u} + y_d H q \bar{d} + y_e H \ell \bar{e}. 
\end{align}
The Boltzmann equations for the asymmetries are
\begin{align}
\dot{n}_q & =  \alpha_3 y_u^2 T \left( - \frac{n_q}{6} - \frac{n_{\bar{u}}}{3} + \frac{n_H}{4} \right) + \alpha_3 y_d^2 T \left( - \frac{n_q}{6} - \frac{n_{\bar{d}}}{3} -  \frac{n_H}{4} \right) \nonumber \\
& \hspace{0.15 in} +  3 \frac{\Gamma_{\rm ws}}{T^3} \Big(-n_q - n_\ell -  \frac{c_W}{3} \dot{\theta} T^2  \Big) +  2 \frac{\Gamma_{\rm ss}}{T^3} \left( -   n_q -n_{\bar{u}}-n_{\bar{d}} - \frac{1}{2} \dot{\theta} T^2 \right), \nonumber \\
\dot{n}_{\bar{u}} & = \alpha_3 y_u^2 T \left( - \frac{n_q}{6} - \frac{n_{\bar{u}}}{3} + \frac{n_H}{4} \right) + \frac{\Gamma_{\rm ss}}{T^3} \left( -  n_q -n_{\bar{u}}-n_{\bar{d}} - \frac{1}{2} \dot{\theta} T^2 \right), \nonumber \\
\dot{n}_{\bar{d}} & = \alpha_3 y_d^2 T  \left( - \frac{n_q}{6} - \frac{n_{\bar{d}}}{3} -  \frac{n_H}{4} \right) + \frac{\Gamma_{\rm ss}}{T^3} \left( -  n_q -n_{\bar{u}}-n_{\bar{d}} - \frac{1}{2} \dot{\theta} T^2 \right), \\
\dot{n}_\ell & =  \alpha_2 y_e^2 T \left( - \frac{n_\ell}{2} - n_{\bar{e}} -  \frac{n_H}{4} \right) +  \frac{\Gamma_{\rm ws}}{T^3} \Big(- n_q - n_\ell -  \frac{c_W}{3} \dot{\theta} T^2 \Big), \nonumber \\
\dot{n}_{\bar{e}} & =   \alpha_2 y_e^2 T \left( - \frac{n_\ell}{2} - n_{\bar{e}} -  \frac{n_H}{4} \right), \nonumber \\
\dot{n}_H & =  - \alpha_3 y_u^2 T \left( - \frac{n_q}{6} - \frac{n_{\bar{u}}}{3} + \frac{n_H}{4} \right) + \alpha_3  y_d^2 T  \left( - \frac{n_q}{6} - \frac{n_{\bar{d}}}{3} -  \frac{n_H}{4} \right) +  \alpha_2 y_e^2 T \left( - \frac{n_\ell}{2} - n_{\bar{e}} -  \frac{n_H}{4} \right),\nonumber \\
\dot{n}_P & = \frac{\Gamma_{\rm ss}}{T^3} \left( - n_q -n_{\bar{u}}-n_{\bar{d}} - \frac{1}{2} \dot{\theta} T^2 \right) + c_W \frac{\Gamma_{\rm ws}}{T^3} \Big(-n_q - n_\ell -  \frac{c_W}{3} \dot{\theta} T^2  \Big), \nonumber
\end{align}
where $c_W$ is the weak anomaly coefficient of the PQ symmetry relative to that of the QCD anomaly.
The total hypercharge and the $B-L$ charge must vanish,
\begin{align}
\label{eq:conserved}
\frac{1}{6} n_q - \frac{2}{3} n_{\bar{u}} + \frac{1}{3}n_{\bar{d}} - \frac{1}{2} n_\ell + n_{\bar{e}} - \frac{1}{2} n_H = 0, \\
\frac{1}{3}(n_q - n_{\bar{u}} - n_{\bar{d}}) -  n_\ell + n_{\bar{e}} =0. \nonumber
\end{align}
Since the scattering by the Yukawa couplings and the sphaleron transition are efficient, the system quickly reaches a quasi-equilibrium state with $\dot{n}_q=\dot{n}_{\bar{u}} =\dot{n}_{\bar{d}}=\dot{n}_\ell=\dot{n}_{\bar{e}}=\dot{n}_H =0$. Further imposing Eq.~(\ref{eq:conserved}), we find that the baryon asymmetry is given by
\begin{align}
\label{eq:nB1gen}
\frac{n_B}{T^3} =
\begin{cases}
\frac{27-32 c_W}{210} \frac{\dot{\theta}}{T} & y_u^2 \ll y_d^2, \alpha_3^3  \\ 
\frac{21-32c_W}{210} \frac{\dot{\theta}}{T} & y_d^2 \ll y_u^2, \alpha_3^3
\end{cases},
\hspace{0.5in} \left(\frac{n_B}{T^3} \ll \frac{2 \, \dot{\theta} \vpq^2}{T^3} \right).
\end{align}
Again, most of the PQ charge is retained in the rotation of the PQ symmetry breaking field. 
In the Standard Model with three generations, whether $y_u^2 \ll y_d^2$ or $y_u^2 \gg y_d^2$ depends on generations. The precise answer in the Standard Model is expected to lie between the two limiting cases in Eq.~(\ref{eq:nB1gen}).
This derivation shows that $c_B = \mathcal{O}(0.1\mathchar`-1)$ unless $c_W \gg 1$, where $c_B$ is defined in the Letter.

%%%%%%%%%%%%%%%%%%%%%%%%%%%
\section{Early Electroweak Phase Transition}
\label{sec:EWPT}
%%%%%%%%%%%%%%%%%%%%%%%%%%%%

\begin{figure}[b]
	\includegraphics[width=0.3\linewidth]{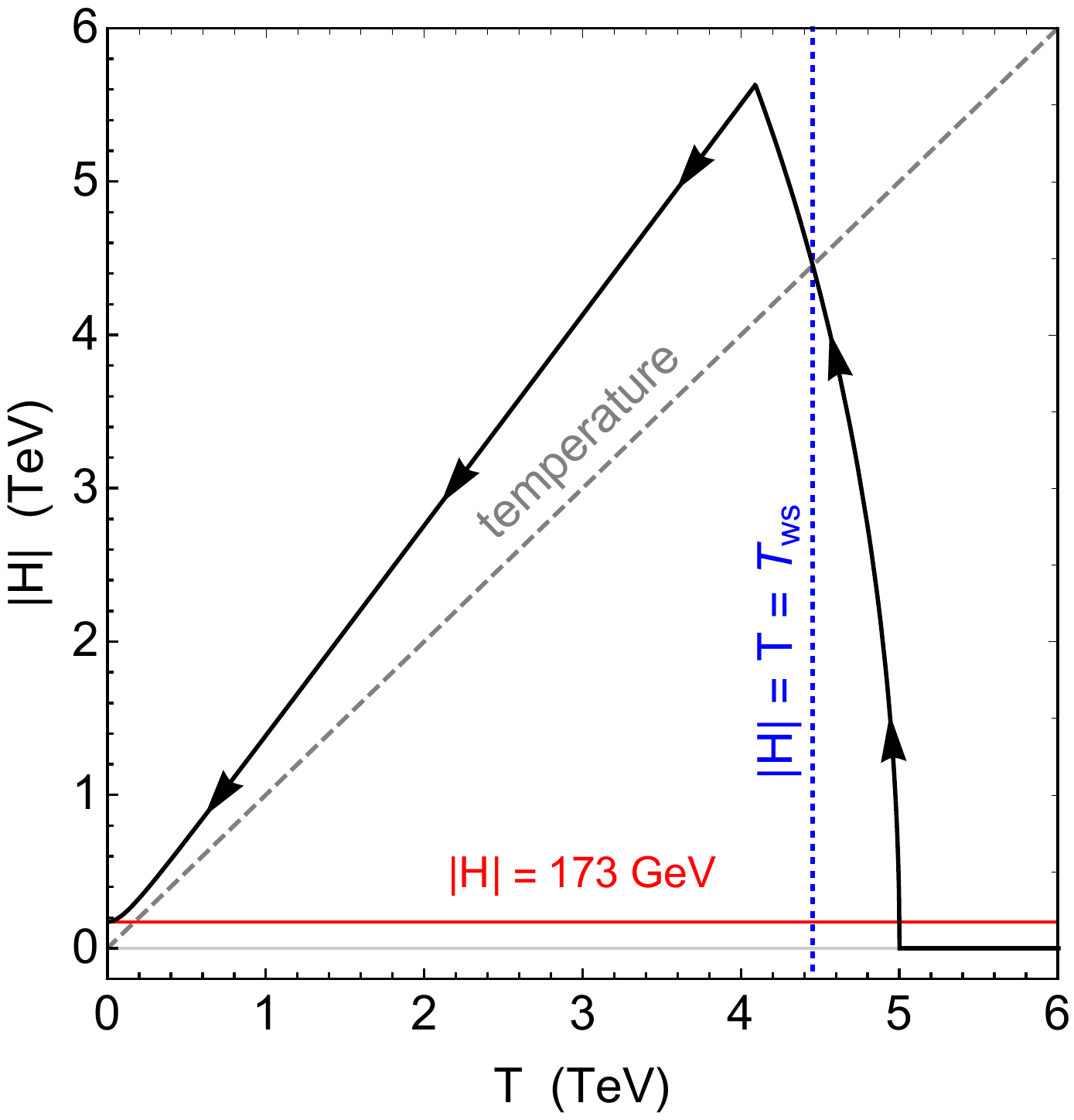}
	\includegraphics[width=0.308\linewidth]{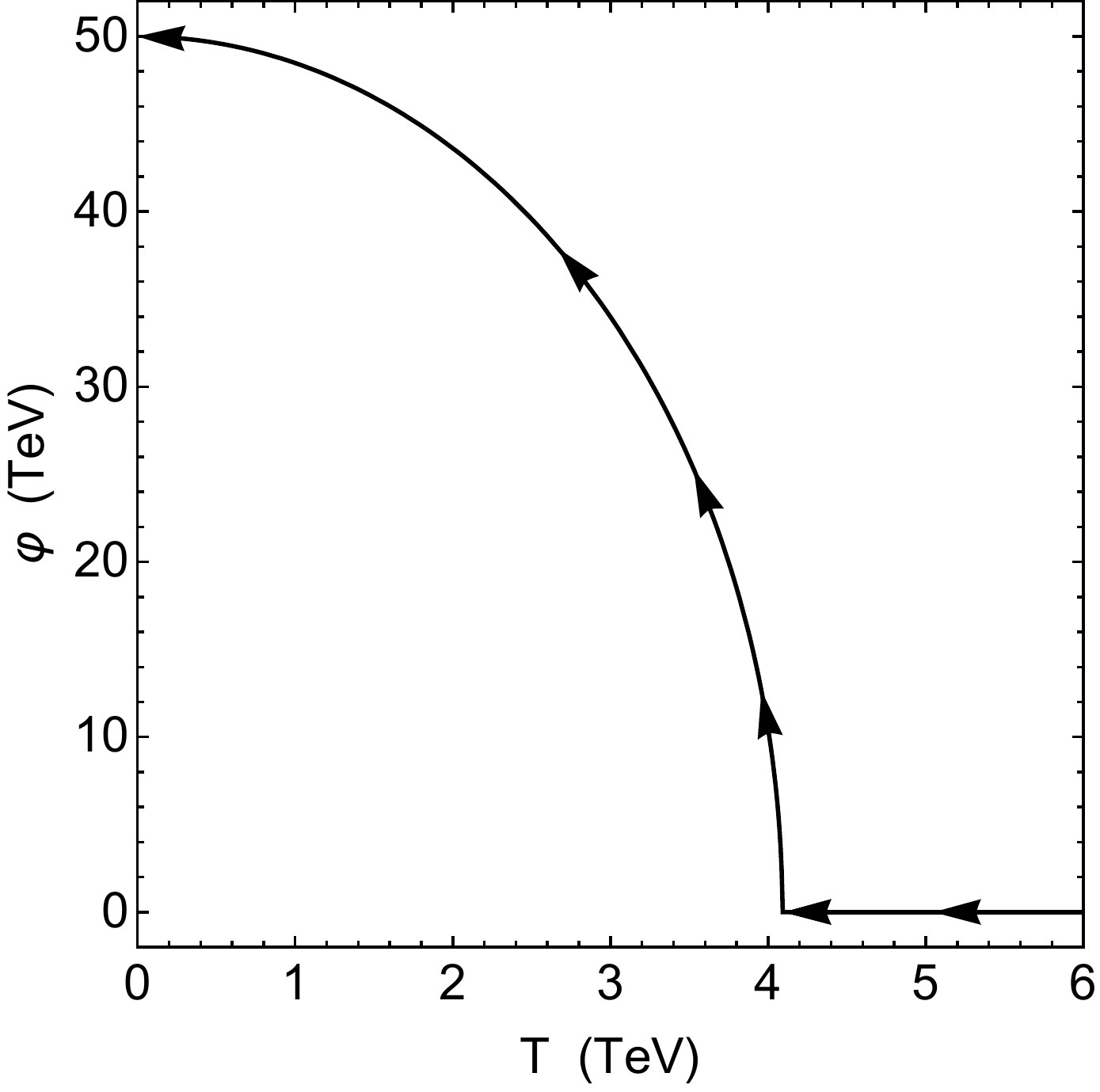}
	\includegraphics[width=0.308\linewidth]{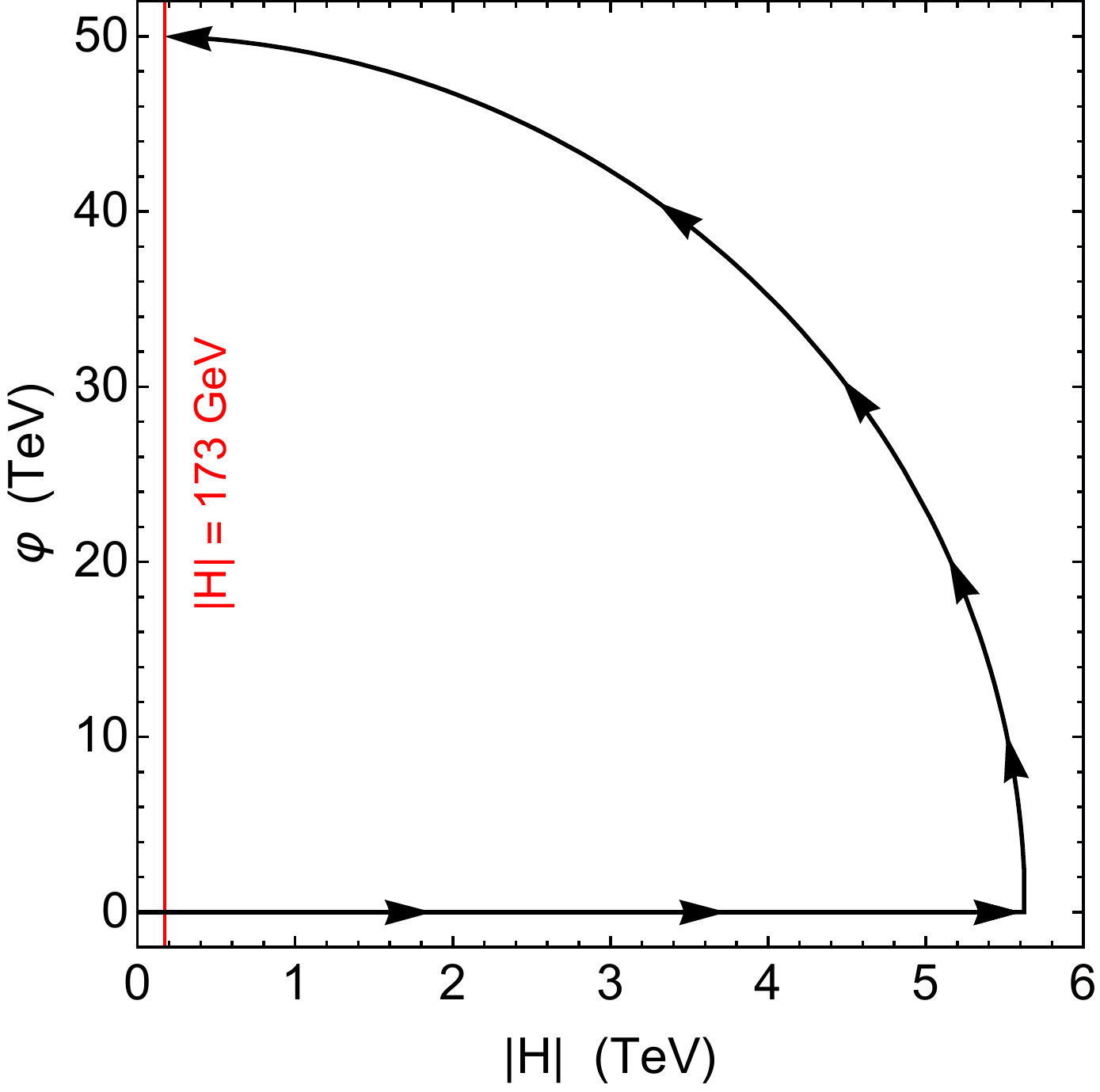}
	\caption{Evolution of scalar fields with $v_S = 50 \TeV, \, v = 173 \GeV, \, \lambda_H = 0.36, \, \kappa = 0.03, \, \lambda = 0.1, \, c_H = 1, {\rm and} \, c_S = 0.5$.}
	\label{fig:H_phi_T}
\end{figure}

We consider the situation where the Higgs $H$ couples to a scalar field $\varphi$ which obtains a vacuum expectation value. We assume that the mass scale appearing in the potential is larger than the electroweak scale, and the electroweak scale appears due to fine-tuning of the parameters. This is the case, for example, with the Next-to-Minimal Supersymmetric Standard Model~\cite{Fayet:1974pd, Nilles:1982dy, Frere:1983ag} with soft masses at the TeV scale or higher. The soft masses above the electroweak scale are required to explain the observed Higgs mass~\cite{Okada:1990gg,Okada:1990vk,Ellis:1990nz,Haber:1990aw} as well as to satisfy the constraints from superpartner searches at the LHC~\cite{Aad:2019ftg,Sirunyan:2019ctn}, so a mild fine-tuning is a generic requirement for supersymmetric theories.

In this situation, the field values of $H$ and $\varphi$ can evolve in the following way. In the early universe with a high temperature, both $H$ and $\varphi$ are trapped at the origin by a thermal potential. As the Universe cools down to a temperature around the mass scale of the potential, $H$ and $\varphi$ develop nonzero field values. At this point, the field value of $\varphi$ is different from the vacuum expectation value, and the quadratic term as well as the field value of $H$ are around the natural scale of the theory. As the Universe cools further, the field value of $\varphi$ gradually approaches the vacuum expectation value. Then the quadratic term of $H$ approaches the electroweak scale, and $H$ eventually reaches the vacuum expectation value.

Here is a concrete example of the potential of $H$ and $\varphi$. Assuming a $Z_2$ symmetry $\varphi \rightarrow -\varphi$, the generic renormalizable potential and the thermal mass terms are parametrized as
\begin{align}
V(H,\varphi) = \lambda_H^2 \left(|H^2|- v^2 \right)^2 +\kappa^2\left( \varphi^2 - v_\varphi^2 \right)^2  + \lambda^2\left(\varphi^2 - v_\varphi^2\right)\left(|H|^2 - v^2\right) + c_H T^2 |H|^2 + c_\varphi T^2 \varphi^2.
\end{align}
At temperatures sufficiently high, the large thermal masses of $H$ and $\varphi$ stabilize these scalars to the origin and the masses of $\varphi$ and $H$ are given by 
\begin{align}
m_H^2 (T) & = T^2 c_H^2 - 2 \lambda_H^2 v^2 - \lambda ^2 v_\varphi^2 , \\
m_\varphi^2 (T) & = T^2 c_\varphi^2 - 2 \kappa ^2 v_\varphi^2 - \lambda ^2 v^2 , \nonumber
\end{align}
where we take $v_\varphi \gg v \simeq 173 \GeV$.
As temperature drops, the thermal mass squared of either $H$ or $\varphi$ first falls below the vacuum mass. Starting at $T = \max(T_H, T_\varphi)$ with
\begin{align}
T_H  = \frac{\sqrt{2 v^2 \lambda _H^2+\lambda ^2 v_\varphi^2}}{c_H} , \hspace{0.5 in}
T_\varphi  = \frac{\sqrt{2 \kappa ^2 v_\varphi^2+\lambda ^2 v^2}}{c_\varphi},
\end{align}
a nonzero expectation value of $|H|$ (or $\varphi$) starts to develop and reaches $\mathcal{O}(\lambda \, v_\varphi / \lambda_H)$ quickly, while the other field is still stabilized at the origin by the thermal mass. Without loss of generality, we assume $T_H > T_\varphi$ to simplify the following discussion. At this point, the electroweak symmetry is broken. When the Higgs expectation value exceeds the temperature, the electroweak sphaleron process falls out of thermal equilibrium at a temperature just below $\lambda \, v_\varphi / c_H$. The nonzero field value of $|H|$ now induces a positive mass squared $\lambda^2 |H|^2$ to $\varphi$ and we require that the negative vacuum mass squared $m_\varphi^2 (T=0) \simeq - 2 \kappa ^2 v_\varphi^2$ still dominates and allows $\varphi$ to obtain a vacuum expectation value eventually. Therefore, when $m_\varphi^2 (T)$ turns negative, a nonzero field value of $\varphi$ starts to develop as well. Since the Higgs vacuum mass of $m_h = 125 \GeV$ is obtained from fine-tuning, the contribution from $\varphi$, i.e.~$\lambda^2 \varphi^2$, gradually cancels the Higgs bare mass $- 2 \lambda_H^2 v^2 - \lambda ^2 v_\varphi^2$ until the Higgs mass at the zero temperature is reached. In other words, the Higgs field value starts to decrease due to the additional mass contribution when $\varphi$ becomes nonzero. As long as the Higgs field value stays larger than the temperature, the electroweak sphaleron process stays out of equilibrium and thus this example has succeeded in raising $T_{\rm ws}$ to a value far above the weak scale.

We explicitly demonstrate the evolution of this potential in Fig.~\ref{fig:H_phi_T} with a fiducial set of parameters listed in the caption. The first two panels show the field values of $|H|$ and $\varphi$ as functions of temperature, whereas the right panel shows the evolution of the potential minimum in the field space. The arrows point in the direction of increasing time and thus decreasing temperature. In the left panel, we observe that electroweak symmetry is broken at $T = T_{H} \simeq 5 \TeV$ while the electroweak sphaleron is decoupled at $T_{\rm ws} \simeq 4.5 \TeV$. As shown in the middle panel, $\varphi$ starts to roll away from the origin at $T \simeq 4 \TeV$. From this temperature, an increase in $\varphi$ causes the Higgs field to decrease, as can be seen in the left panel. This evolution ends when $\varphi$ settles to the minimum $v_\varphi$ so that the fine-tuned Higgs mass and vacuum expectation value are obtained. Since $|H|$ stays larger than temperature after $T_{\rm ws}$, the electroweak sphaleron processes are kept out of thermal equilibrium. Larger values of $T_{\rm ws}$ than assumed here can be easily obtained by larger values of $v_\varphi$. Hence, this concrete example realizes the assumptions made in the main analysis, where $T_{\rm ws} \gg 100 \GeV$.

%%%%%%%%%%%%%%%%%%%%%%%%%%%
\section{A Supersymmetric Realization of Axiogenesis}
\label{sec:SUSY_axio}
%%%%%%%%%%%%%%%%%%%%%%%%%%%%

In this section we investigate a supersymmetric realization of axiogenesis.
We assume that the PQ symmetry is explicitly broken by a higher dimensional operator in the superpotential,
\begin{equation}
\label{eq:explicitPQB}
W = \frac{P^{d+1}}{M^{d-2}},
\end{equation}
where $P$ is the PQ symmetry breaking field and $M$ is a mass scale. This explicit breaking can generate the PQ asymmetry.
(It is also possible to initiate the axion rotation by another PQ charged scalar such as the heavy KSVZ squarks or the Higgs doublets in the DFSZ model. This may break the electroweak symmetry, which should be restored in the early universe so that the conversion of the PQ asymmetry into the $B+L$ asymmetry occurs.)

We require that $P$ has a flat PQ invariant potential in order to obtain a large enough condensate and the PQ asymmetry. Examples include 1) a model with the PQ symmetry breaking by dimensional transmutation due to the running of the soft mass~\cite{Moxhay:1984am}, 
\begin{equation}
V = m^2 |P|^2 \left( {\rm ln}\frac{|P|^2}{\vpq^2} -1  \right),
\end{equation}
2) a two-field model with soft masses,
\begin{equation}
W = X( P \bar{P}- \vpq^2), ~~V_{\rm soft} = m_P^2 |P|^2 + m_{\bar{P}}^2 |\bar{P}|^2,
\end{equation}
where $X$ is a chiral multiplet whose $F$-term fixes the PQ symmetry breaking fields $P$ and $\bar{P}$ along the moduli space $P \bar{P} = \vpq^2$,
and 3) PQ symmetry breaking by quantum corrections in gauge mediation~\cite{ArkaniHamed:1998kj,Asaka:1998ns,Asaka:1998xa}. Then in the early universe, $P$ may receive a sufficiently negative Hubble induced mass and obtain a large initial field value. The explicit PQ symmetry breaking in Eq.~(\ref{eq:explicitPQB}) is then effective, causing a potential and thus a motion in the angular direction.

To be concrete we assume that the potential of the radial direction of $P$, called the saxion $S$ with mass $m_S$, is well approximated by a quadratic potential at large field values. We also assume a negative Hubble induced mass of $P$. Then the PQ symmetry breaking field value is determined by the balance between the Hubble induced mass term and the $F$-term potential given by the superpotential in Eq.~(\ref{eq:explicitPQB})~\cite{Dine:1995kz,Harigaya:2015hha},
\begin{equation}
\label{eq:V_S}
V(S) \sim - H^2 S^2 +  \frac{S^{2d}}{M^{2d-4}}.
\end{equation}
When $S_i$ is large, the saxion mass is mainly given by a gravity-mediated mass $m_{S, g}$, which is comparable to the gravitino mass $m_{3/2}$. 
As the Hubble scale $H$ drops below the mass $m_{S,g}$, the PQ symmetry breaking field begins oscillating with an initial saxion field value $S_i$
\begin{equation}
S_i \sim \left( m_{S, g}^2 M^{2d-4}\right)^{\frac{1}{2d-2}}.
\end{equation}
The explicit PQ symmetry breaking given by the $A$-term associated with the superpotential in Eq.~(\ref{eq:explicitPQB}),
\begin{equation}
V(P) \sim d \, m_{3/2} \frac{P^{d+1}}{M^{d-2}} + {\rm h.c.},
\end{equation}
is not negligible. Hence, while the saxion starts to oscillate, explicit breaking kicks $P$ toward the angular direction with a speed of order $m_{3/2}$. The asymmetry of the PQ charge given by the angular motion is, at the beginning of the oscillation,
\begin{equation}
n_{{\rm PQ}, i} = \dot{\theta}_i f_i^2 \sim \frac{2}{N_{\rm DW}} \, m_{3/2} S_i^2 ,
\end{equation}
where the angular misalignment from the minimum is assumed to be $\mathcal{O}(1)$ and $N_{\rm DW}$ is the number of domain walls. It is convenient to normalize the asymmetry by the number density of the saxion,
\begin{equation}
\frac{n_{\rm PQ}}{n_S} \equiv \epsilon, \hspace{0.5 in} \epsilon \sim \frac{4}{N_{\rm DW}} \frac{m_{3/2}}{m_{S, g}},
\end{equation}
because this is a redshift-invariant quantity. The scaling of $n_{\rm PQ} \propto a^{-3}$ can be understood as a result of PQ charge conservation.
The parameter $\epsilon$ is expected to be order unity and is treated as a free parameter in what follows. 
(We assume that the potential of the radial direction is not shallower than the quadratic one, so that the possible instability to form solitons~\cite{Coleman:1985ki,Kusenko:1997si,Enqvist:1997si,Enqvist:1998en,Kasuya:1999wu} is absent.)

The energy density of $P$ must be depleted eventually by thermalization in order to avoid overclosure. After $P$ is thermalized, only the energy density of the radial mode is depleted, while the energy density associated with the rotation remains. This is because it is free-energetically favorable to keep most of the charge in the form of rotation rather than particle excitations. This energy density $\rho_{\rm PQ}$ scales the same way as matter, $\rho_{\rm PQ} \propto a^{-3}$, when the saxion field value $\veff \gg \vpq$ and then scales as kination, $\rho_{\rm PQ} \propto a^{-6}$, when $\veff \simeq \vpq$. Therefore, $\rho_{\rm PQ}$ simply redshifts away and no further depletion mechanism is needed. Rigorous discussions of the dynamics are given in the previous part of this Supplemental Material.

In the remaining of this section, we derive the expression for the baryon asymmetry. The PQ charge stored in the axion rotation is converted into $B+L$ by the QCD and electroweak sphaleron transitions. Due to the large initial condensate of the PQ breaking field $P$, the energy density at low temperatures tends to be dominated by that of $P$, which we will assume in what follows. 

From the onset of the $P$ oscillation until thermalization at temperature $T_{\rm th}$, the PQ charge number density $n_{\rm PQ}$ and the number density of the radial mode $n_{\rm S}$ scale the same way. After the radial mode is depleted to create a thermal bath with a temperature $T_{\rm th}$, the yield of the PQ asymmetry is a constant given by
\begin{equation}
Y_{\rm PQ} \equiv \frac{n_{\rm PQ}}{s} = \epsilon \frac{3 T_{\rm th}}{4 m_S} ,
\end{equation}
which, with $n_{\rm PQ} =  \dot{\theta} f_a^2$, implies that the angular speed is
\begin{equation}
\dot{\theta}(T) = \epsilon \frac{g_* \pi^2 }{30} \frac{T_{\rm th}T^3}{m_S \feff^2(T)},
\end{equation}
where $\feff (T)$ is the effective axion decay constant at temperature $T$, i.e.~$\sqrt{2} \left| P(T) \right|/N_{\rm DW}$ with $P(T)$ the field value of $P$ at $T$.
Using the yield of the baryon asymmetry,
\begin{equation}
\label{eq:YB_thetadot}
Y_B = \frac{n_B}{s} = \left. \frac{45 c_B}{2 g_* \pi^2} \frac{\dot{\theta}}{T} \right|_{T = T_{\rm ws}} \hspace{-0.2 in},
\end{equation}
as presented in the Letter, we obtain
\begin{equation}
\label{eq:YB}
Y_B =  \epsilon  \frac{ 3 c_B T_{\rm th} T_{\rm ws}^2 }{4 m_S \feff^2(T_{\rm ws})} .
\end{equation}
This expression is valid whether thermalization or the electroweak phase transition occurs first. 

\begin{figure}
	\includegraphics[width= 0.5\linewidth]{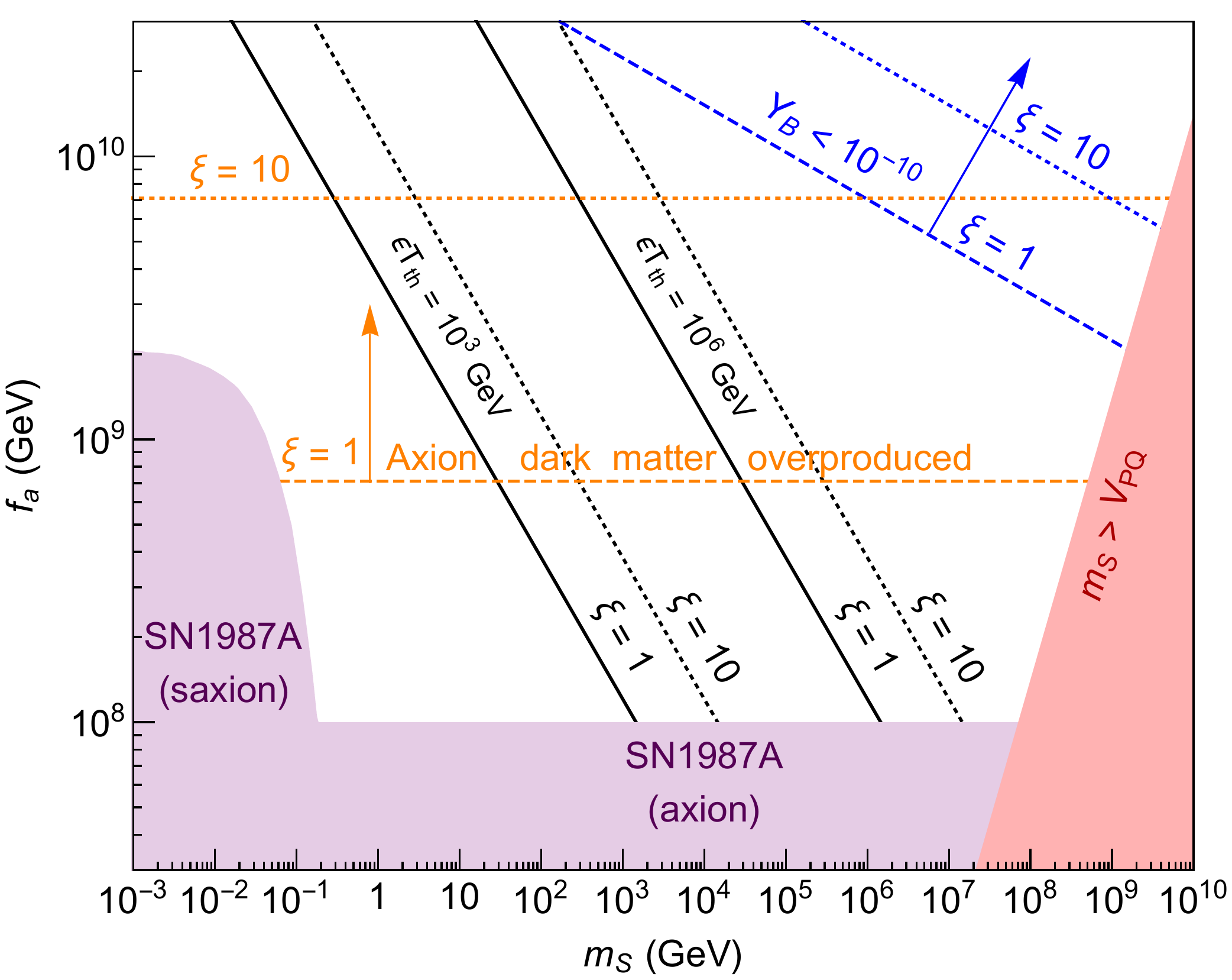}
	\caption{The parameter space compatible with the observed baryon asymmetry.}
	\label{fig:fa_ms_SM}	
\end{figure}

We demonstrate the viable parameter space in Fig.~\ref{fig:fa_ms_SM}
with $N_{\rm DW} = 1$.
Based on the discussion in the Letter, the region above the orange line is excluded due to axion dark matter overproduction for $\xi = 1$ (dashed) and $\xi = 10$ (dotted), where 
\begin{equation}
\xi \equiv \left( \frac{c_B}{100} \right) \left( \frac{T_{\rm ws} }{ 130 \GeV} \right)^2 .
\end{equation}

We first discuss the case where $T_{\rm th} > T_{\rm ws}$. One can use conservation of PQ charge and energy to determine $\feff^2(T)$ for $T  < T_{\rm th}$
\begin{equation}
\label{eq:Veff_T}
\feff^2(T) = \max \left[ f_a^2, \ \ \epsilon \frac{g_* \pi^2}{30 N_{\rm DW}} \frac{ T_{\rm th}^4 }{m_S^2} \left(\frac{T}{T_{\rm th}}\right)^3 \right] ,
\end{equation}
where the former (latter) corresponds to the case where the saxion has (not) relaxed to $\vpq$ at $T$. This leads to
\begin{equation}
\label{eq:YB_th}
Y_B = \min \left[ \epsilon \frac{ 3 c_B T_{\rm th} T_{\rm ws}^2 }{4 m_S f_a^2} , \ \ \frac{45 N_{\rm DW} c_B}{2 g_* \pi^2} \frac{m_S}{T_{\rm ws}} \right] ,
\end{equation}
which implies a maximum achievable amount of baryon asymmetry and sets a lower bound on $m_S$
\begin{equation}
\label{eq:ms_bound}
m_S >  \frac{2 g_* \pi^2 Y_B T_{\rm ws}}{45 N_{\rm DW} c_B }  \simeq \frac{0.2 \MeV}{N_{\rm DW}} \left( \frac{T_{\rm ws}}{5 \TeV} \right) \left( \frac{0.1}{c_B} \right).
\end{equation}
The black lines show the required values of $T_{\rm th}$ determined by the first case in Eq.~(\ref{eq:YB_th}), assuming $T_{\rm th} > T_{\rm ws}$.

For $T_{\rm th} < T_{\rm ws}$,
although Eq.~(\ref{eq:YB}) still applies, the determination of $\veff(T_{\rm ws})$ depends on the thermalization process of the saxion because the temperature and Hubble relationship changes during the matter-dominated era with entropy injection. Despite this model dependence, there always exists $T_{\rm th}$ in the parameter space of interest that reproduces the observed $Y_B$. Specifically, one can show that the required thermalization temperature
\begin{equation}
T_{\rm th} \simeq \frac{45 \MeV}{\epsilon}  \left( \frac{\feff(T_{\rm ws})}{ 10^9 \GeV} \right)^2 \left(\frac{m_S}{\rm MeV}\right) \left( \frac{5 \TeV}{ T_{\rm ws} } \right)^2 \left( \frac{0.1}{c_B} \right)
\end{equation}
is always larger than $\mathcal{O}(\rm MeV)$ and does not affect Big Bang nucleosynthesis.

We now discuss the thermalization channels necessary to realize the required values of $T_{\rm th}$. The radial mode can scatter with thermal particles via renormalizable couplings, e.g.~$ W \supset y P Q \bar{Q}$, or high dimensional operators such as $W \supset P^2 H_u H_d / M$. In both cases, the scattering rate $\Gamma_{\rm th}$ at $T_{\rm th}$ has an upper bound
\begin{equation}
\label{eq:Gamma_th}
\Gamma_{\rm th} (T_{\rm th}) \sim N_{\rm th} \times \begin{cases}
y^2 T_{\rm th} \\
\frac{V_{\rm eff}^2(T_{\rm th}) T_{\rm th}^3 }{M^2}
\end{cases} \hspace{-0.1in}
 \lesssim \frac{N_{\rm th} T_{\rm th}^3}{\veff^2(T_{\rm th})}, 
\end{equation}
when one ensures that the particles the saxion couples to (assuming $N_{\rm th}$ of them) are in the thermal bath as assumed, i.e.~$y V_{\rm eff}(T_{\rm th}) < T_{\rm th}$ or $V_{\rm eff}^2(T_{\rm th}) /M < T_{\rm th}$.
Together with Eq.~(\ref{eq:Veff_T}) in the high $T_{\rm th}$ limit and the thermalization condition $H(T_{\rm th}) = \Gamma_{\rm th} (T_{\rm th})$, Eq.~(\ref{eq:Gamma_th}) implies
\begin{equation}
T_{\rm th} \lesssim 10^{11} \GeV \left( \frac{N_{\rm th}}{\epsilon \, N_{\rm DW}} \right)^{ \scalebox{1.01}{$\frac{1}{3}$} } \left(\frac{m_S}{10^8 \GeV}\right)^{ \scalebox{1.01}{$\frac{2}{3}$} } .
\end{equation} 
The region above the blue line in Fig.~\ref{fig:fa_ms_SM} is excluded with $\xi = 1$ (dashed) and $\xi = 10$ (dotted) because $T_{\rm th}$ necessary for baryogenesis exceeds such an upper bound. In other words, the finite scattering rate cannot consistently realize the high thermalization temperatures required.  In computing the blue lines, we have assumed $\epsilon = N_{\rm th} = 1$ and the upper bound of $f_a \propto \epsilon^{1/3} N_{\rm th}^{1/6}$ is rather insensitive to the choice of $\epsilon$ and $N_{\rm th}$.

It is assumed that the saxion dominates the energy density of the Universe, in which case the above result is insensitive to the initial condition of the saxion. This assumption holds as long as the initial field value of the saxion is sufficiently large
\begin{equation}
S_i \gtrsim 4 \times 10^{16} \GeV \left(\frac{\rm TeV}{m_S}\right)^{ \scalebox{1.01}{$\frac{1}{4}$} }  \left(\frac{T_{\rm th}}{10^6 \GeV}\right)^{ \scalebox{1.01}{$\frac{1}{2}$} }  ,
\end{equation}
which is easily achievable based on Eq.~(\ref{eq:V_S}).

In the red region, the radial mode mass $m_S$ exceeds the unitarity limit.
The purple region is excluded since the emission of saxions or axions in a supernova core affects the duration of the neutrino emission~\cite{Ellis:1987pk,Raffelt:1987yt,Turner:1987by,Mayle:1987as,Ishizuka:1989ts,Raffelt:2006cw,Chang:2018rso,Carenza:2019pxu}. The constraint from the saxion emission can be evaded by introducing a large enough saxion-Higgs mixing to trap saxions inside the core. The large mixing can be achieved in the DFSZ model. In summary, Fig.~\ref{fig:fa_ms_SM} shows that a wide range of the saxion mass $m_S$ is viable, while a low $f_a$ is favored in the minimal realization of axiogenesis. 

\end{document}